\documentclass[a4paper,11pt]{article}

\usepackage{lipsum}
\usepackage{multicol}

\newcommand{\surf}{\mathcal{S}}

\newcommand{\R}{\mathbb{R}}

\newcommand{\normal}{\boldsymbol{\nu}}

\newcommand{\vel}{\boldsymbol{u}}

\newcommand{\relvel}{\boldsymbol{v}}

\newcommand{\param}{\boldsymbol{X}}
\newcommand{\paramC}{X}

\newcommand{\coord}{\boldsymbol{x}}

\newcommand{\shop}{\boldsymbol{B}}

\newcommand{\meanc}{\mathcal{H}}

\newcommand{\proj}{\boldsymbol{P}}
\newcommand{\Id}{\boldsymbol{I}}

\newcommand{\inputTikzPic}[1]{\ifthenelse{\boolean{plotTikzPics}}{\input{#1}}{\fbox{\centering\begin{minipage}[t][5cm][t]{0.9\textwidth}\centering\textbf{\color{red}enable plotTikzPics}\end{minipage}}}}

\newcommand{\Grad}{\nabla}
\newcommand{\Div}{\operatorname{div}}%
\newcommand{\DivSurf}{\Div_{\!\surf}}%
\newcommand{\GradSurf}{\Grad_{\!\surf}}
\newcommand{\GradP}{\Grad_{\!\proj}}
\newcommand{\GradC}{\Grad_{\!C}}

\newcommand{\DivP}{\operatorname{div}_{\!\proj}\!}
\newcommand{\DivC}{\operatorname{div}_{\!C}\!}

\newcommand{\pressure}{p}

\newcommand{\Be}{\textup{Be}}

\newcommand{\ReynoldsNumber}{\mathrm{Re}}
\newcommand{\BendingNumber}{\mathrm{Be}}

\renewcommand{\Re}{\ReynoldsNumber}
\renewcommand{\Be}{\BendingNumber}

\newcommand{\stressC}{\sigma}
\newcommand{\stress}{\boldsymbol{\stressC}}

\usepackage{graphicx}
\usepackage{dcolumn}
\usepackage{bm}
\usepackage{amsmath}
\usepackage{amssymb}
\usepackage{mathtools}
\usepackage{epstopdf, epsfig}
\usepackage{color}
\usepackage{tikz}
\usepackage{pgfplots}
\usepackage{float}
\usepackage[a4paper,left=2.5cm,right=2.5cm]{geometry}
\usepackage{hyperref}
\hypersetup{
    colorlinks = true,
    urlcolor   = blue,
    citecolor  = black,
}

\begin{document}

\title{Wrinkling of fluid deformable surfaces}

\author{
  Veit Krause$^1$, Axel Voigt$^{1,2,3}$\\[1em]
  $^{1}${\small Institute of Scientific Computing, TU Dresden, 01062 Dresden, Germany}\\
  $^{2}${\small Center of Systems Biology Dresden, Pfotenhauerstr. 108, 01307 Dresden, Germany}\\
  $^{3}${\small Cluster of Excellence, Physics of Life, TU Dresden, Arnoldstr. 18, 01307 Dresden, Germany}
}



\maketitle

\begin{abstract}
Wrinkling instabilities of thin elastic sheets can be used to generate periodic structures over a wide range of length scales. Viscosity of the thin elastic sheet or its surrounding medium has been shown to be responsible for dynamic processes. While this has been explored for solid as well as liquid thin elastic sheets we here consider wrinkling of fluid deformable surfaces, which show a solid-fluid duality and have been established as model systems for biomembranes and cellular sheets. We use this hydrodynamic theory and numerically explore the formation of wrinkles and their coarsening, either by a continuous reduction of the enclosed volume or the continuous increase of the surface area. Both lead to almost identical results for wrinkle formation and the coarsening process, for which a universal scaling law for the wavenumber is obtained for a broad range of surface viscosity and rate of change of volume or area. However, for large Reynolds numbers and small changes in volume or area wrinkling can be suppressed and surface hydrodynamics allows for global shape changes following the minimal energy configurations of the Helfrich energy for corresponding reduced volumes.
\end{abstract}



\section{Introduction}

Thin elastic sheets that resist bending, are ubiquitous in nature and man-made systems and their mechanical properties enable countless applications. At a critical compression such sheets wrinkle. Thereby wrinkles emerge with a wavelength which balances the resistance to bending of the sheet and the stiffness that resists out-of-plane deformations. This might lead to the formation of periodic structures with regular spacing. The required compression can be induced by different approaches. Recent studies have focused on understanding the bending deformations that occur when solid thin elastic sheets are stretched \cite{huang2007,Davidovitch2011,nayyar2014stretch}, poked \cite{vella2011wrinkling, vella2015indentation}, or wrapped around a curved object \cite{king2012elastic, hure2012stamping}. Under suitable conditions, wrinkles can also be dynamic, with a tunable wavelength \cite{box2019dynamics}. This can be realized if the solid thin elastic sheet is floating on a liquid-air interface. Wrinkles are no longer static, they coarsen and thereby change wavelength and amplitude. The coarsening of the wrinkles has been associated with the inertial response of the underlying liquid. In contrast to these results for solid thin elastic sheets, wrinkling instabilities have also been observed in liquid thin elastic sheets. Besides wrinkling at the collapse of a viscous bubble film upon rupture \cite{debregeas1998life,silveira2000rippling,savva2009viscous,le2012buckling,oratis2020new}, they can also form in a controlled manner \cite{oratis2020new}. As bending forces are no longer present, such sheets buckle when the rate of compression is faster than the smoothing effect of surface tension and wrinkling is associated with the viscosity of the film. While the mechanisms for wrinkling of solid and liquid thin elastic sheets significantly differ, both dynamics are associated with viscosity.

Various theories have been developed to model solid thin elastic sheets. Among them the F\"oppl-von K\'arm\'an equation is a popular approach \cite{Davidovitch2011,puntel2011wrinkling,healey2013wrinkling,taylor2014spatial}. For a mathematical approach, which interprets wrinkling of solid sheets as a geometrical problem and determines the wrinkling patterns as configurations with the least possible energy, we refer to \cite{tobasco2022exact}. In contrast with this, approaches for liquid thin elastic sheets are rare. In \cite{oratis2020new} a dynamic version of the F\"oppl-von K\'arm\'an equation is used. Even, if argued in \cite{oratis2020new} that inertia plays a critical role in determining the number of wrinkles, hydrodynamic theories have not been considered in this context. This also holds for theoretical approaches for the dynamic wrinkling considered in \cite{box2019dynamics}.

We provide such a hydrodynamic approach and focus on fluid deformable surfaces. Such surfaces can be seen as thin elastic sheets with solid and liquid properties and have been established as model systems for biomembranes \cite{art:Arroyo2009,art:torres-sanchez_millan_arroyo_2019}. They exhibit a solid-fluid duality, while they store elastic energy when stretched or bent, as solid thin elastic sheets, they flow as viscous two-dimensional fluids under in-plane shear. With these properties, fluid deformable surfaces can be seen as a hybrid between solid and liquid thin elastic sheets. Depending on the material properties, solid-like behaviour (large viscosity) or liquid-like behaviour (low bending rigidity) can be achieved. The solid-fluid duality has several consequences: it establishes a tight interplay between tangential flow and surface deformation. In the presence of curvature, any shape change is accompanied by a tangential flow and, vice-versa, the surface deforms due to tangential flow. Wrinkling of lipidic membranes have, e.g., been observed in \cite{C9SM01902B}. Besides biomembranes, fluid deformable surfaces are also appropriate to model the complex deformations in embryonic epithelia. These cellular sheets are fluid-like to enable morphogenetic flows, e.g. measured during vertebrate body axis elongation \cite{mongera2018fluid}, and wrinkling is an essential process in morphogenetic tissue deformations.

Models describing this interplay between curvature and surface flow have been introduced and numerically solved in \cite{art:torres-sanchez_millan_arroyo_2019,reuther2020numerical,art:Krause22,bachini2023derivation}. These numerical approaches provide the basis to explore the impact of surface viscosity and bending rigidity on the dynamics of wrinkles in biomembranes or epithelial tissue. However, instead of such concrete applications, we consider a systematic approach, where wrinkles form in a controlled manner and their dynamics can be analyzed. Similar to  \cite{oratis2020new} we consider a closed surface with an embedded volume and either dynamically reduce the volume or increase the surface area to induce the required compression resulting in wrinkling. By considering a prolate geometry we further provide an axis of symmetry to induce wrinkling in a deterministic way. We also neglect the potential influence of the bulk fluid phases. This is justified if the Saffman-Delbr\"uck number 
\cite{SD_PNAS_1975}, which defines a hydrodynamic length relating the viscosity of the elastic sheet and the surrounding bulk fluid, is large. In this case, the hydrodynamics is effectively 2D on spatial scales smaller than the Saffman-Delbr\"uck number \cite{PhysRevE.81.011905}. Within this approach, we computationally study the effect of surface viscosity, respectively the Reynolds number, and the rate of volume reduction or area extension on the formation of wrinkles and their coarsening. 

The remainder of this paper is structured as follows: In Section \ref{sec:methods} we introduce the mathematical model and briefly describe its numerical solution. In Section \ref{sec:results} we describe the computational setup and vary parameters to explore the dynamics of wrinkling comparing the two different approaches of reducing the volume and increasing the surface area. This is considered within a geometric setting which allows for a controlled wrinkling instability. In Section \ref{sec:discussion} we discuss the results, demonstrate a universal scaling law for the wavenumber, generalize the setting to a less controlled situation and draw conclusions. Detailed computations and validations of the numerical approach are provided in Appendix \ref{app:sec:app1}.

\section{Methods} \label{sec:methods}
\subsection{Notation}
We consider a time-dependent smooth domain $\Omega=\Omega(t)\subset\R^3$ and its boundary $\surf = \surf(t)$, which is given by a parametrization $\param$. Related to $\surf$ we denote the surface normal $\normal$, the shape operator $\shop= -\GradP \normal$, the mean curvature $\meanc= \operatorname{tr}\shop$ and the surface projection $\proj=\Id-\normal\otimes\normal$, with $\operatorname{tr}$ the trace operator and $\Id$ the identity. 
Let $\vel$ be a continuously differentiable $\R^3$-vector field and $\stress$ a continuously differentiable $\R^{3\times3}$-tensor field defined on $\surf$. As in \cite{jankuhn2018,Nitschke2022GoI,bachini2023derivation} we define the surface tangential gradient $\GradP\vel = \proj (\nabla\vel^e)_{|\surf} \proj$ and the componentwise surface gradient $\GradC\stress = (\nabla\stress^e)_{|\surf} \proj$ where $\vel^e$ and $\stress^e$ are arbitrary smooth extensions of $\vel$ and $\stress$ in the normal direction and $\nabla$ is the gradient of the embedding space $\R^3$. For the relation to the corresponding covariant derivative $\GradSurf$, we refer to \cite{bachini2023derivation}. We define the corresponding divergence operators for a vector field $\vel$ and a tensor field $\stress \proj$ by $\DivP\vel=\operatorname{tr}(\GradP\vel)$ and $\DivC (\stress\proj) = \operatorname{tr}(\GradC \stress)$. With this definition $\DivC \stress$ leads to a non-tangential vector field even if $\stress$ is a tangential tensor field. By $\Delta_{C} = \DivC \GradC$ we denote the componentwise Laplacian, with the Laplace-Beltrami operator $\Delta_\surf = \DivSurf \GradSurf$ acting on each component. 

\subsection{Model}
In \cite{art:Krause22} a model for fluid deformable surfaces with conserved enclosed volume is considered. Here, this model is modified by allowing for changes in volume and area, now reading
\begin{align} 
  \label{eq:navierstokes}
  \begin{aligned}
    \partial_t \vel + \nabla_{\relvel}\vel &= -\GradSurf p-p\meanc\normal + \frac{2}{\Re}\DivC \stress -\gamma \vel \\
    & \quad - \lambda\normal + \mathbf{b}  \\
    \DivP \vel &= \delta_A  \\
    \int_\surf \vel\cdot\normal &= -\delta_V
  \end{aligned}    
\end{align}
where $\vel$ is the surface velocity, $p$ is the surface pressure, $\stress=\frac{1}{2}(\GradP \vel + \GradP \vel^T)$ is the rate of deformation tensor, $\mathbf{b} = \frac{1}{\Be}(-\Delta_\surf\meanc -\meanc(\Vert\shop\Vert^2 - \frac{1}{2}\operatorname{tr}(\shop)^2)\normal$ is the bending force, $\lambda$ is the Lagrange multiplier to ensure the volume constraint, and $\relvel = \vel-\partial_t\param$ is the relative material velocity. $\Re$ denotes the Reynolds number, $\Be$ is the bending capillary number and $\gamma$ is a friction coefficient. In \cite{art:Krause22} the enclosed volume and the surface area are considered to be constant, thus $\delta_V = \delta_A = 0$. Here we consider these volume and surface area terms time-dependent and define 
\begin{align}
\delta_V(t) &= P_V\cdot(1+\tanh(\alpha(t-t_0))), \\
\delta_A(t) &= P_A\cdot(1+\tanh(\alpha(t-t_0))),
\end{align}
where $t_0>0$ is a relaxation time, $P_V$ is the volume reduction rate, $P_A$ is the area growth rate and $\alpha \gg 1$ is a parameter ensuring a smooth transition. Within the time interval $0<t<t_0$ this leads to $\delta_V\approx 0$ and $\delta_A\approx 0$ and for $t>t_0$ we obtain $\delta_V\approx P_V$ and $\delta_A\approx P_A$. This results in a change in volume $\frac{d}{dt}\vert\Omega\vert = P_V$ and surface area $\frac{d}{dt}\vert S\vert = P_A \vert \surf \vert$, respectively. While a reduction in volume directly corresponds to the experiments in \cite{oratis2020new}, growth in surface area introduces a new mechanism. In the context of fluid deformable surfaces it has been introduced in \cite{toshniwal2021isogeometric}. In this context, $\delta_A$ considers a mass production term. It allows to model growth or swelling of the surface.

We will relate $\delta_A$ and $\delta_V$ to ensure the same reduced volume. The reduced volume is defined as the scaled quotient of volume and surface area by
\begin{align} \label{eq:Vr}
V_r &= 6 \sqrt{\pi} \frac{\vert\Omega\vert}{\vert\surf\vert^{\frac{3}{2}}},
\end{align}
see \cite{Seifert_AP_1997}. For $P_A = \frac{2}{3}P_V / \vert\Omega\vert$ we thus obtain the same reduced volume $V_r$, see Appendix \ref{app:sec:app1} for the detailed computation.

We thus obtain a model for a fluid deformable surface with dynamic change of the reduced volume, either by reducing the enclosed volume or by increasing the surface area.

\subsection{Numerics}
For the numerical discretization, we follow the concept of \cite{art:Krause22} and consider an Eulerian-Lagrangian approach. Lagrangian in normal directions with 
\begin{align}
        \partial_t \param \cdot \normal &= \vel\cdot \normal \label{eq:param}
\end{align}
and Eulerian in tangential direction. To deal with potential mesh distortions we combine equations \ref{eq:navierstokes} and \ref{eq:param} with the mesh redistribution approach introduced in \cite{barrett2008parametric}, considering 
\begin{align}
    \meanc \normal &= \Delta_{C} \param \label{eq:meanc}
\end{align}
which in addition provides a representation of the mean curvature $\meanc$.

We approximate the surface $\surf$ by a third-order polynomial surface $\surf_h$. The equations \ref{eq:navierstokes}-\ref{eq:meanc} and all geometrical quantities like the normal vector $\normal_h$ and the shape operator $\shop_h$ are formulated with respect to $\surf_h$. To solve the equations numerically, we discretize the system in time by a semi-implicit Euler method and in space by a surface finite element (SFEM) approach \cite{dziuk2013finite,nestler2019finite}. For the pair of velocity and pressure, we use the Taylor-Hood element (third-order velocity $\vel_h$ and second-order pressure $\pressure_h$) and also consider a third-order approximation for the mean curvature $\meanc_h$. For a parallel efficient treatment of $\lambda_h$, we refer to \cite{art:Krause22}. The discrete equations are implemented in the finite element toolbox AMDiS \cite{vey2007amdis,witkowski2015software} which is based on DUNE \cite{SanderDune2020,alkamper2014dune}. For the surface approximation, we used the Dune-CurvedGrid library  \cite{praetorius2020dunecurvedgrid}. For a straightforward mesh parallelization and multiple processor computation, we use the PETSc library and solve the linear system by a direct solver. Pre- and postprocessing are done in Paraview and Python.  

For validation of the numerical model for $\delta_V = \delta_A = 0$ we refer to \cite{art:Krause22}. Additional validation for $\delta_V \neq 0$ and $\delta_A \neq 0$ are considered in Appendix \ref{app:sec:app1}. 

\begin{figure}[t]
    \centering
    \includegraphics[width=\textwidth]{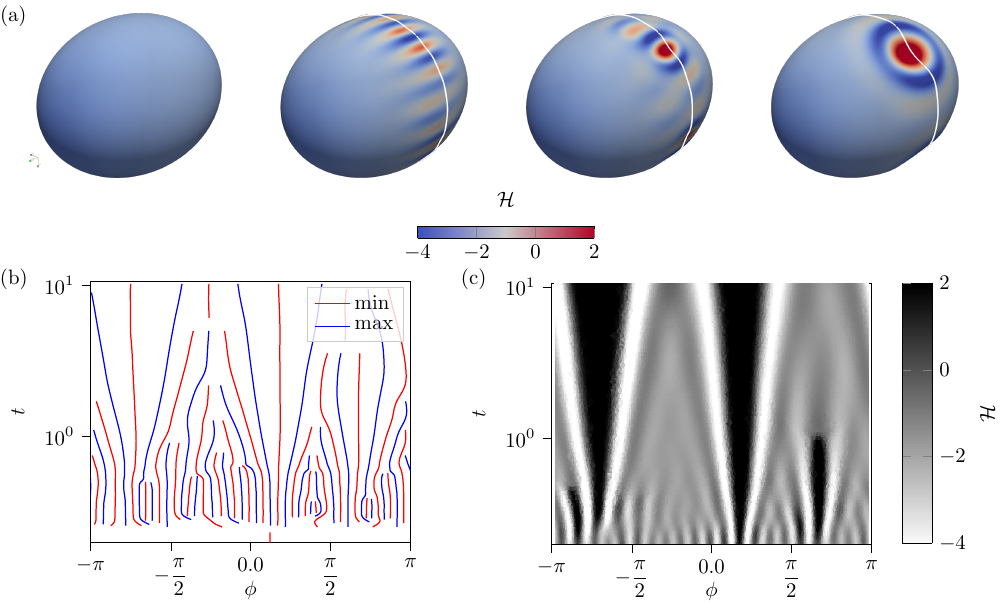}    
    \caption{Wrinkling obtained by volume reduction with $P_V = 0.02$ and $\Re = 0.025$. (a) Snapshots for $t=0.0,0.4,0.6,1.2$ colour coded by  mean curvature $\meanc$. The wrinkles are analysed along the equator denoted by the white line. (b) Minima and maxima of wrinkle profile along the rotational angle of the equator over time. (c) Mean curvature $\meanc$ of wrinkle profile along the rotational angle of the equator over time. 
    }
    \label{fig:process}
\end{figure} 

\section{Results} \label{sec:results}

We consider a prolate geometry with volume $\vert \Omega \vert = 4.19$ and surface area $\vert \surf \vert = 12.73$, slightly perturbed (see Appendix \ref{app:sec:app1}), without flow $\vel = \mathbf{0}$, as initial condition. This setting leads to a reduced volume $V_r = 0.98$. The minimal energy configuration of the Helfrich energy corresponds to a dumbbell shape and remains within this regime during the time evolution with reducing volume or increasing surface area \cite{Seifert_AP_1997}. Figure \ref{fig:process} (a) shows a typical evolution. It is obtained by volume reduction with $P_V = 0.02$ and Reynolds number $\Re = 0.025$. Wrinkles form around the equator of the prolate and are oriented in direction of the smaller principal curvature of the surface. They are regularly arranged during formation and relax over time. Figure \ref{fig:process} (b), (c) visualize the coarsening process by considering the wrinkle profile along the equator, indicated by the white line in Figure \ref{fig:process} (a). 

We will explore these dynamics by varying the Reynolds number $\Re$ and the volume reduction rate $P_V$ or the area growth rate $P_A$. We consider low Reynolds numbers $\Re \leq 1$ in a range which can be related to the experiments in \cite{oratis2020new,silveira2000rippling}. Instead of an instantaneous compression, as, e.g., considered in \cite{box2019dynamics,PhysRevFluids.2.014202} we consider finite rates of volume reduction and area increase. We are especially interested in the formation of wrinkles and their coarsening and will therefore analyze the number of wrinkles and their wavelength over time. All other parameters remain constant. We consider $\Be = 0.01$, $\gamma = 0.0$, $\alpha = 50$ and $t_0 = 0.1$. 

\begin{figure}[t]
    \centering
    \includegraphics[width=\textwidth]{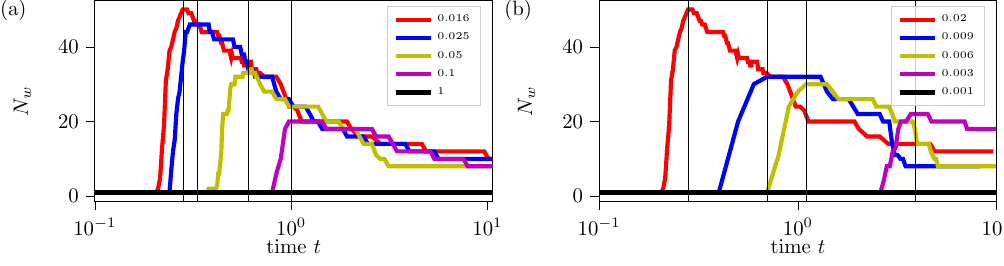}
    \caption{Number of wrinkles $N_w$ over time for continuous volume reduction. (a) Considered for different Reynolds number $\Re$ (for $P_V=0.02$) and (b) for different volume reduction rates $P_V$ (for $\Re=0.016$).}
    \label{fig:viscosity}
\end{figure}

\subsection{Wrinkling by volume reduction}

We consider $\delta_A = 0$. In Figure \ref{fig:viscosity} (a) we show the number of wrinkles $N_w$ over time for $P_V = 0.02$ and different Reynolds numbers $\Re$. Depending on $\Re$ a threshold of compression has to be overcome before wrinkles form. If this threshold is reached the maximal number of wrinkles $N_{max} = \max N_w(t)$ forms almost instantaneously. The time instances at which the maximal number of wrinkles $N_{max}$ forms are highlighted in Figure \ref{fig:viscosity} (a). The time of wrinkle formation $\tau_0 = \operatorname{argmin} \{ t \vert N_w(t) > 0\}$ increases with increasing $\Re$. This increase is associated with a decrease of $N_{max}$. For $\Re = 1$, no wrinkling occurs. In this situation tangential flow is sufficient to globally deform $\surf$ to suppress wrinkling. After wrinkles are formed they coarsen. The coarsening process is almost independent on $\Re$, the curves for different $\Re$ in Figure \ref{fig:viscosity} (a) essentially lye on top of each other.

We now vary $P_V$ and keep $\Re = 0.016$. The corresponding results for the number of wrinkles $N_w$ 
are shown in Figure \ref{fig:viscosity} (b). We again mark the time instances of wrinkle formation $\tau_0$. For low volume reduction rates, no wrinkling occurs. Again in this situation, tangential flow is sufficient to globally deform $\surf$ to suppress wrinkling. If wrinkling occurs, qualitatively the dependency on $P_V$ is similar to the dependency on $1/\Re$. The wrinkle formation for a faster volume reduction corresponds to that of a lower Reynolds number. 

\begin{figure}[t]
    \centering   
    \includegraphics[width=\textwidth]{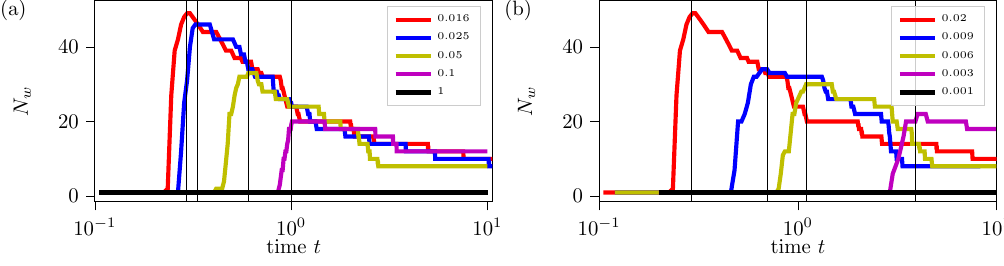}
    \caption{Number of wrinkles $N_w$ over time for continuous area increase. (a) Considered for different Reynolds number $\Re$ (for $P_A$ corresponding to $P_V=0.02$) and (b) for different increasing area rates $P_A$ (for $\Re=0.016$). Instead of $P_A$, we indicate the corresponding values for $P_V$ for better comparison.}
    \label{fig:viscosity2}
\end{figure}

\begin{figure}[t]
    \centering
    \includegraphics[width = \textwidth]{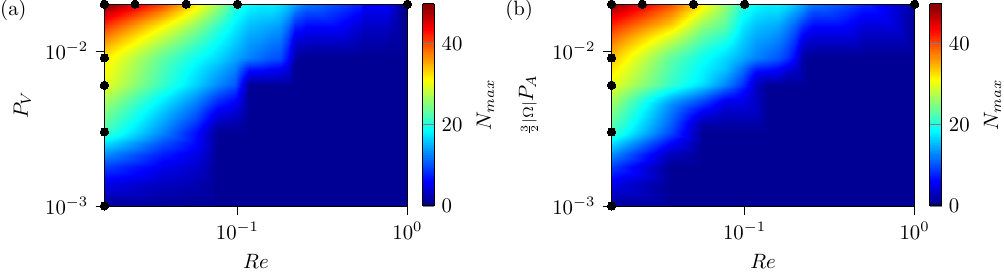} 
    \caption{Phase diagram of the maximal number of wrinkles $N_{max}$ over the Reynolds number $\Re$ and the volume reduction rate $P_V$ (a) and the area increase rate $P_A$ (b). Instead of $P_A$ we indicate $\frac{3}{2}\vert\Omega\vert P_A$, which are the corresponding values for $P_V$, for better comparison. The black dots highlight values of $\Re$ and $P_V$ shown in Figures \ref{fig:viscosity} or \ref{fig:viscosity2}, respectively.}
    \label{fig:phasediag}
\end{figure}

Due to the continuous reduction of volume, the surface does not relax to the corresponding equilibrium shape associated with the reduced volume (dumbbell shape) \cite{Seifert_AP_1997} if the Reynolds number $\Re$ is too small or the volume reduction rate too large. The induced wrinkling along the short axis of the prolate and their subsequent coarsening lead to an energy path towards a local minimum, similar to the one shown in Figure \ref{fig:process} (a) at $t = 1.2$. Decreasing the volume further can easily be realized by deepening the remaining valleys. This further increases the energy barriers, which need to be overcome to escape this local minimum. While the kinetic energy globally in time decreases during the coarsening process, with only local in time increasing periods, which are associated with coarsening events, the Helfrich energy strongly increases for all $\Re$ with only small perturbations associated with the coarsening events, see Figure \ref{app:fig:energies} (a) and (c) in Appendix \ref{app:sec:app1}, respectively. However, for $\Re = 1$ or above and small volume reduction rates the tangential flow allows for global shape changes continuously sampling shapes close to the equilibrium dumbbell shapes \cite{Seifert_AP_1997}, see Figure \ref{app:fig:energies} (a) in Appendix \ref{app:sec:app1}. This corresponds to previous results on enhanced convergence towards equilibrium shapes due to surface viscosity \cite{krause2023surface}. The equilibrium shapes are computed by minimizing the Helfrich energy under the constraints of volume and global area conservation following \cite{krause2023surface,olshanskii2023,bachini2023derivation}. Additional qualitative differences can also be seen in the tangential flow field, which is plotted for different Reynolds numbers $\Re$ in Figure \ref{app:fig:flow} 
in Appendix \ref{app:sec:app1}. While wrinkles induce perturbations of the flow field around the equator, for $\Re = 1$ tangential flow is only used for mass transport to globally deform the surface towards a more elongated dumbbell shape.

\subsection{Wrinkling by area increase}

We next ask if the behaviour changes if instead of a volume reduction the surface area is increased. In analogy to Figure \ref{fig:viscosity} (a) and (b) we show in Figure \ref{fig:viscosity2} (a) and (b) the number of wrinkles over time for different Reynolds number $\Re$ (for $\frac{3}{2} \vert \Omega\vert P_A = 0.02$) and for different area increasing rates $P_A$ ($\Re = 0.016$). The results are very similar to the corresponding plots in Figure \ref{fig:viscosity}.

\subsection{Phase diagram}

\begin{figure}[t]
    \centering
    \includegraphics[width=\textwidth]{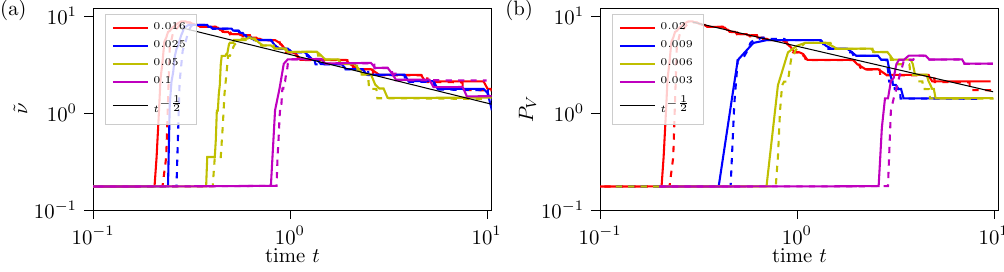} 
    \caption{Coarsening of wrinkles analysed by the wavenumber $\tilde \nu$ as a function of time. (a) Considered for different Reynolds numbers $\Re$ and $P_V=0.02$ (solid lines) and corresponding $P_A$ (dashed line). (b) Considered for different volume reduction rates $P_V$ (solid lines) and corresponding area increase rates $P_A$ (dashed lines) and $\Re=0.016$. The data correspond to Figures \ref{fig:viscosity} and \ref{fig:viscosity2}, but only those values which lead to wrinkling are considered. Both indicate a scaling law of $t^{-1/2}$ indicated by the black lines.}
    \label{fig:conclusion}
\end{figure}

In Figure \ref{fig:phasediag} the maximal number of wrinkles $N_{max}$ is shown with respect to the Reynolds number $\Re$ and the volume reduction rate $P_V$ or the corresponding rescaled area increasing rate $P_A$. The qualitative behaviour of the dynamics is the same as in Figures \ref{fig:viscosity} (a) and (b) and \ref{fig:viscosity2} (a) and (b), we therefore only focus on $N_{max}$. The specific values from Figures \ref{fig:viscosity} (a) and (b) and \ref{fig:viscosity2} (a) and (b) are marked for comparison. While slight quantitative differences can be observed, the qualitative behaviour remains and does not depend on the realization of the reduced volume by volume reduction or area increase.

\section{Discussion} \label{sec:discussion}

Fluid deformable surfaces with an enclosed volume and a continuous reduction of volume or a continuous increase in surface area can undergo a wrinkling instability. Only for large Reynolds numbers $\Re = 1$ or above and for small changes in volume $\delta_V$ or area $\delta_A$ this instability can be suppressed. In these cases, surface viscosity allows to globally deform the surface to adapt to the changing reduced volume $V_r$. The emerging dynamic shapes are close to the equilibrium shapes of the Helfrich energy \cite{Seifert_AP_1997}. Surface hydrodynamics has been shown to support convergence towards these shapes if the volume and area remain constant, $\delta_V = \delta_A = 0$, see \cite{art:Krause22}. This result is extended towards small changes in volume or area. Also under these circumstances surface hydrodynamics allows for global shape adaptation and continuously sampling the equilibrium shapes or at least shapes which are close to them. The evolution of the energies, the kinetic energy $E_{kin}$ and the Helfrich energy $E_H$ is shown in Figure \ref{app:fig:energies} (a) and (c) in Appendix \ref{app:sec:app1}, respectively. The reference values for $E_H$ correspond to the equilibrium shapes for the reduced volume $V_r$. This shows a remarkable difference to the theory in \cite{tobasco2022exact}. Under the presence of viscosity the configuration with the least possible energy does not contain wrinkles. Within the considered geometrical setting the least possible energy corresponds to the equilibrium (dumbbell) shape of the Helfrich energy \cite{Seifert_AP_1997}. Wrinkling under the influence of viscosity only leads to local energy minima. 

The evolution after winkles have formed is characterized by coarsening. This has been qualitatively addressed in Figures \ref{fig:process}, \ref{fig:viscosity} and \ref{fig:viscosity2}. The coarsening process seems to be independent on $\Re$ and also, with slightly larger error bars, variations in $\delta_V$ and $\delta_A$ lead to similar coarsening. We now quantify this. Instead of the number of wrinkles $N_w$ we therefor consider the surface wavenumber 
\begin{align}
  \tilde\nu = \frac{N_w}{l} 
\end{align}
where $l$ is the geodesic length of the curve along which we measure the wrinkles. This curve is highlighted by the white line in Figure \ref{fig:process} (a). The corresponding results to Figures \ref{fig:viscosity} and \ref{fig:viscosity2} are shown in Figure \ref{fig:conclusion} (a) and (b), respectively. They indicate a coarsening rate $t^{-1/2}$. This scaling behaviour is independent on $\Re$ and $P_V$ or $P_A$. This further highlights that wrinkling only depends on the change in the relation of enclosed volume and surface area. The same effect can be achieved by reducing the volume or by growing the surface. This might explain the similar wrinkling pattern found in nature and man-made systems. Within morphogenetic tissue deformations actually both, volume changes and surface growth might be essential. 

\begin{figure}[t]
    \centering
    \includegraphics[width=\textwidth]{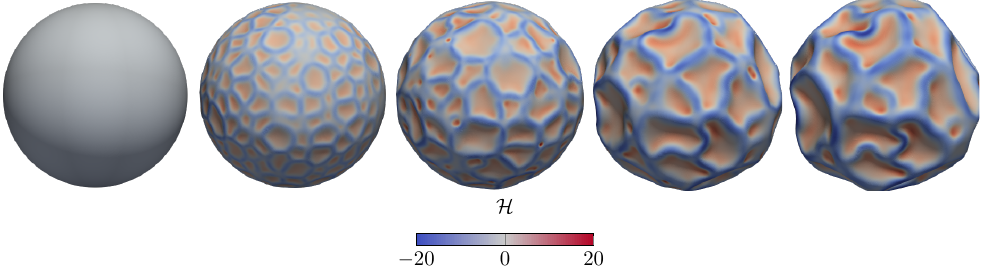}
    \caption{Wrinkling instability and subsequent coarsening on the unit sphere for $\Re=0.016$ and $P_A=1$.0. Time instances are shown for  $t=0.0,0.1,0.2,0.4,0.6$, from left to right. The color coding corresponds to the mean curvature $\meanc$. A corresponding video of the evolution is provided in the Electronic Supplement.}
    \label{fig:sphere}
\end{figure}

A theoretical foundation of the obtained scaling law and also an experimental validation of the influence of viscosity in the proposed model currently have to remain open issues. The considered geometric setting was chosen to initiate wrinkles in a controlled manner. However, the numerical approach also works for other settings. On a sphere the emerging wrinkling patterns are less controlled due to the missing prefered symmetry axis. Considering a spherical domain with significantly larger growth rates leads to symmetry breaking and more irregular surface patterns, see Figure \ref{fig:sphere}. The results correspond to Reynolds number $\Re = 0.016$ and $P_A = 1$. All other parameters are as before. Shown are time instances of the evolution, the color coding corresponds to the mean curvature $\meanc$. A corresponding movie of the evolution is provided in the Electronic Supplement. The emerging wrinkling morphology is reminiscent to wrinkle patterns in swelling films on spherical substrates \cite{C3SM27331H}, which were computational reproduced in \cite{stoop2015curvature} using a phenomenological pattern formation model. Even if dynamics is not considered in these studies, the observed hexagonal wrinkles can also be found in the dynamic patterns under the influence of surface viscosity, see Figure \ref{fig:sphere}. This provides additional similarities between wrinkling patterns in solid and liquid thin 
elastic sheets and asks for experimental validation in the context of biomembranes or epithelial tissue.
\appendix

\section{Appendix}
\label{app:sec:app1}

\subsection{Computations on the reduced volume}
\label{app:reducedVolume}

According to the definition of the reduced volume in eq. \eqref{eq:Vr} we obtain
\begin{align*}
  \frac{\mathrm{d}}{\mathrm{d} t} V_r = \frac{6 \sqrt{\pi}}{\vert \surf \vert^{\frac{3}{2}}} \frac{\mathrm{d}}{\mathrm{d} t} \vert\Omega\vert \!-\! \frac{9 \vert \Omega\vert \sqrt{\pi}}{\vert \surf \vert^{\frac{5}{2}}} \frac{\mathrm{d}}{\mathrm{d} t} \vert\surf\vert = \frac{6 \sqrt{\pi}}{\vert \surf \vert^{\frac{3}{2}}} \delta_V \!-\! \frac{9 \vert \Omega\vert \sqrt{\pi}}{\vert \surf \vert^{\frac{3}{2}}} \delta_A.
\end{align*}
To fulfill $ \frac{\mathrm{d}}{\mathrm{d} t} V_r = 0$ thus requires the chose $P_A = \frac{2}{3} P_V / \vert \Omega \vert$.

\subsection{Construction of the surface}
Let $\param$ be a parametrization of the unit sphere. We define the parametrization of the prolate by $$\param_{r,a} =r((1+a)^{-\frac{1}{2}}\paramC_0,(1+a)^{-\frac{1}{2}}\paramC_1,(1+a)\paramC_2).$$ The elongation is determined by $a$, which is chosen as $a=0.2$. The radius is chosen by $r=1 + \epsilon r_0(\coord)$ where $\epsilon>0$ and $r_0$ is a Gaussian random field 
\begin{align*}
    r_0(\coord) = \sum\limits_{i=0}^N \alpha_i e^{-\beta_i \Vert \coord-\coord_i  \Vert^2 }
\end{align*}
with uniformly distributed variables $\{ \alpha_i\}_i^N\subset [-10^{-6},10^{-6}]$ , $\{ \beta_i \}_i^N\subset [100,1000]$ and $\{ \coord_i \}_i^N\subset\surf$, see \cite{Salvalaglio_2020}. This initial surface guarantees a surface distortion independent of the discretization. For the spherical surface we set $a = 0$.  

\begin{figure}[t]
    \centering
    \includegraphics[width=\textwidth]{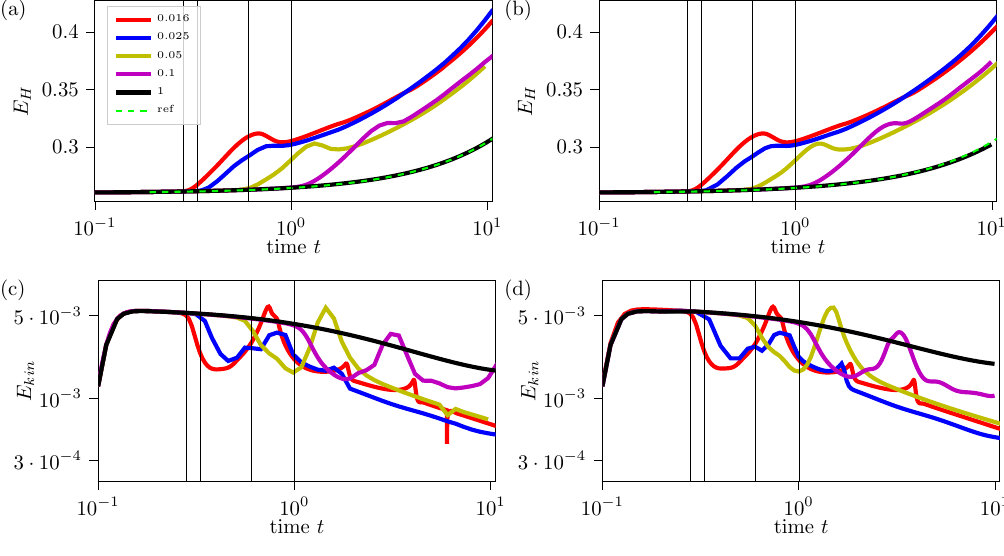}
    \caption{Evolution of the Helfrich energy $E_H$ and the kinetic energy $E_{kin}$ for different Reynolds numbers $\Re$. (a) and (c) for decreasing volume with $P_V = 0.002$ and (b) and (d) for increasing surface area with $P_A = \frac{3}{2}\vert \Omega \vert P_V$. The corresponding plots are almost identical. As a reference, we plotted the minimal Helfrich energy for the corresponding reduced volume (dashed green line).}
    \label{app:fig:energies}
\end{figure}

\subsection{Energies}

Throughout the paper, we consider the non-dimensional energies  
\begin{align*}
    E_{H} &= \frac{1}{\Be} \int_{\surf} \frac{1}{2}\meanc^2, \quad
    E_{kin} = \int_{\surf} \frac{1}{2}\Vert\vel\Vert^2, 
\end{align*}
denoting the Helfrich energy and the kinetic energy, respectively. 

The evolution of these energies corresponding to the simulations shown in Figures \ref{fig:viscosity} and \ref{fig:viscosity2} are shown in Figure \ref{app:fig:energies} (a) and (b), respectively. They indicate a stronger increase in $E_H$ for lower Reynolds numbers $\Re$ and an increase of $E_{kin}$ until the wrinkling instability is followed by a decay which is interrupted by coarsening events.

\begin{figure}[t]
    \centering
    \includegraphics[width=\textwidth]{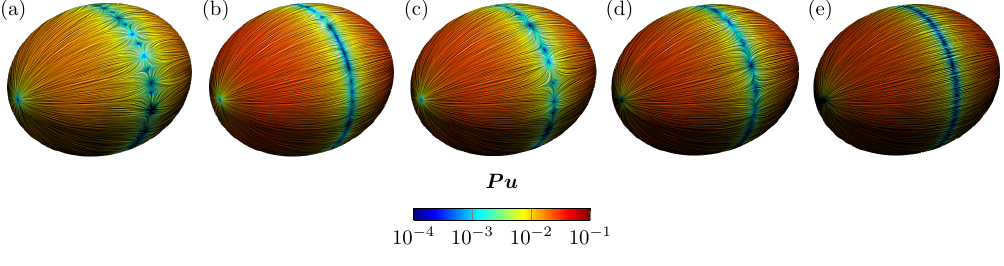}
    \caption{Visualization of the tangential fluid flow by the surface LIC filter. The colour demonstrates the magnitude of the tangential velocity. The plots correspond to the time instants of the maximal number of wrinkles highlighted in Figure \ref{fig:viscosity} (a) for $P_V=0.02$ and different Reynolds number (a) $\Re=0.016$, (b) $\Re=0.025$, (c) $\Re=0.05$, (d) $\Re=0.1$, (e) $\Re=1.0$. }
    \label{app:fig:flow}
\end{figure}

\subsection{Numerical validation}
\label{app:sub:validation}
\begin{figure}[t]
    \centering
    \includegraphics[width=\textwidth]{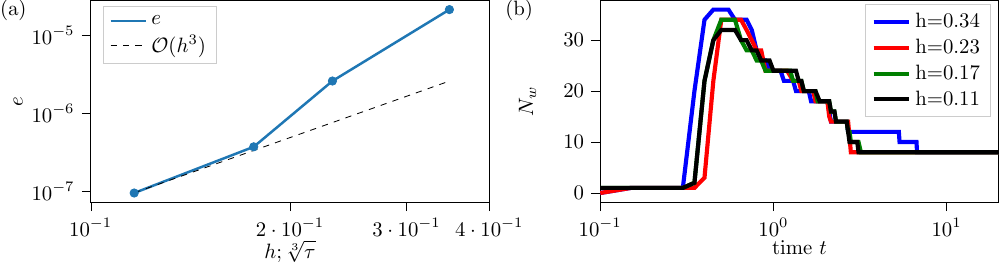}
    \caption{Convergence study for continuous volume reduction. The considered parameters are $\Re = 0.05$ and $P_V = 0.02$. $h$ denotes the mesh size. (a) Inextensibility error $e$ indicates third-order convergence in space and first-order convergence in time. (b) Number of wrinkles over time. The time wrinkling occurs and the coarsening process are almost indistinguishable for fine mesh sizes.}
    \label{fig:validation}
\end{figure}

Numerical analysis results for these types of problems exist only for Stokes flow on stationary surfaces \cite{Hardering2022,HP2023,R2024}. The proposed optimal orders of convergence have been confirmed numerically in \cite{Brandneretal_SIAMJSC_2022} and have also been obtained for fluid deformable surfaces \cite{krause2023surface,bachini2023derivation}. In \cite{krause2023surface} it is shown that the additional volume constraint does not alter the convergence properties. We reproduce the results for the volume reduction case. As before we reduce the time step width with the grid size by $\tau \sim h^3$ and consider the inextensibility error $e = \Vert \DivP \vel_h \Vert_{L^\infty(L^2)}$. As in \cite{krause2023surface} we observe roughly third-order convergence for $e$ with respect to $h$ and first-order with respect to $\tau$. According to \cite{Hardering2022,HP2023,R2024} these are the optimal rates, to be expected for the considered setting. In addition, we consider the number of wrinkles $N_w(t)$ and at least qualitatively obtain the wrinkle instability and the coarsening process to be independent of the discretization, see Figure \ref{fig:validation}.

\paragraph{Ethics} This work did not require ethical approval from a human subject or animal welfare committee.
\paragraph{Data Accessibility} Data is available upon reasonable request.
\paragraph{Autor's Contributions} V.K. and A.V. developed the model and conceived the computational experiments. V.K. implemented the simulation algorithm and conducted the computational experiments. V.K. and A.V. analyzed the results and wrote the manuscript.
\paragraph{Competing Interesrts} We declare we have no competing interests.
\paragraph{Funding} V.K. and A.V. were supported by the German Research Foundation (DFG) within grant FOR3013. We further acknowledge computing resources provided at FZ J\"ulich within grant pfamdis and at ZIH/TU Dresden within grant WIR. 





\bibliography{lib}

\begin{thebibliography}{10}

\bibitem{alkamper2014dune}
M.~Alk{\"a}mper, A.~Dedner, R.~Kl{\"o}fkorn, and M.~Nolte.
\newblock The dune-alugrid module.
\newblock {\em arXiv preprint arXiv:1407.6954}, 2014.

\bibitem{art:Arroyo2009}
M.~Arroyo and A.~DeSimone.
\newblock Relaxation dynamics of fluid membranes.
\newblock {\em Phys. Rev. E}, 79:031915, 2009.

\bibitem{bachini2023derivation}
E.~Bachini, V.~Krause, I.~Nitschke, and A.~Voigt.
\newblock Derivation and simulation of a two-phase fluid deformable surface
  model.
\newblock {\em J. Fluid Mech.}, 977:A41, 2023.

\bibitem{barrett2008parametric}
John~W Barrett, Harald Garcke, and Robert N{\"u}rnberg.
\newblock On the parametric finite element approximation of evolving
  hypersurfaces in $\mathbb{R}^3$.
\newblock {\em J. Comput. Phys.}, 227(9):4281--4307, 2008.

\bibitem{box2019dynamics}
Finn Box, Doireann O’Kiely, Ousmane Kodio, Maxime Inizan, Alfonso~A
  Castrej{\'o}n-Pita, and Dominic Vella.
\newblock Dynamics of wrinkling in ultrathin elastic sheets.
\newblock {\em Proc. Natl. Acad. Sci. USA}, 116:20875--20880, 2019.

\bibitem{Brandneretal_SIAMJSC_2022}
P.~Brandner, T.~Jankuhn, S.~Praetorius, A.~Reusken, and A.~Voigt.
\newblock Finite element discretization methods for velocity-pressure and
  stream function formulations of surface {S}tokes equations.
\newblock {\em SIAM J. Sci. Comput.}, 44:A1807--A1832, 2022.

\bibitem{C3SM27331H}
Derek Breid and Alfred~J. Crosby.
\newblock Curvature-controlled wrinkle morphologies.
\newblock {\em Soft Matter}, 9:3624--3630, 2013.

\bibitem{Davidovitch2011}
Benny Davidovitch, Robert~D. Schroll, Dominic Vella, Mokhtar Adda-Bedia, and
  Enrique~A. Cerda.
\newblock Prototypical model for tensional wrinkling in thin sheets.
\newblock {\em Proc. Natl. Acad. Sci. USA}, 108(45):18227--18232, 2011.

\bibitem{debregeas1998life}
G~d Debr{\'e}geas, P-G De~Gennes, and F~Brochard-Wyart.
\newblock The life and death of" bare" viscous bubbles.
\newblock {\em Science}, 279(5357):1704--1707, 1998.

\bibitem{dziuk2013finite}
Gerhard Dziuk and Charles~M Elliott.
\newblock Finite element methods for surface pdes.
\newblock {\em Acta Numer.}, 22:289--396, 2013.

\bibitem{Hardering2022}
H.~Hardering and S.~Praetorius.
\newblock Tangential errors of tensor surface finite elements.
\newblock {\em IMA J. Num. Anal.}, 43:1543–1585, 2022.

\bibitem{HP2023}
Hanne Hardering and Simon Praetorius.
\newblock A parametric finite-element discretization of the surface stokes
  equations.
\newblock {\em arXiv preprint arXiv:2309.00931}, 2023.

\bibitem{healey2013wrinkling}
Timothy~J Healey, Qingdu Li, and Ron-Bin Cheng.
\newblock Wrinkling behavior of highly stretched rectangular elastic films via
  parametric global bifurcation.
\newblock {\em J Nonlinear Sci}, 23:777--805, 2013.

\bibitem{PhysRevE.81.011905}
M.~L. Henle and A.~J. Levine.
\newblock Hydrodynamics in curved membranes: The effect of geometry on
  particulate mobility.
\newblock {\em Phys. Rev. E}, 81:011905, 2010.

\bibitem{huang2007}
Jiangshui Huang, Megan Juszkiewicz, Wim Jeu, Enrique Cerda, Todd Emrick,
  Narayanan Menon, and Thomas Russell.
\newblock Capillary wrinkling of floating thin polymer films.
\newblock {\em Science}, 317(5838):650--653, 2007.

\bibitem{hure2012stamping}
J{\'e}r{\'e}my Hure, Beno{\^\i}t Roman, and Jos{\'e} Bico.
\newblock Stamping and wrinkling of elastic plates.
\newblock {\em Phys. Rev. Lett.}, 109(5):054302, 2012.

\bibitem{jankuhn2018}
Thomas Jankuhn, Maxim~A. Olshanskii, and Arnold Reusken.
\newblock {Incompressible fluid problems on embedded surfaces: Modeling and
  variational formulations}.
\newblock {\em {Interfaces Free Boundaries}}, {20}({3}):{353--377}, {2018}.

\bibitem{king2012elastic}
Hunter King, Robert~D Schroll, Benny Davidovitch, and Narayanan Menon.
\newblock Elastic sheet on a liquid drop reveals wrinkling and crumpling as
  distinct symmetry-breaking instabilities.
\newblock {\em Proc. Natl. Acad. Sci. USA}, 109(25):9716--9720, 2012.

\bibitem{PhysRevFluids.2.014202}
Ousmane Kodio, Ian~M. Griffiths, and Dominic Vella.
\newblock Lubricated wrinkles: Imposed constraints affect the dynamics of
  wrinkle coarsening.
\newblock {\em Phys. Rev. Fluids}, 2:014202, Jan 2017.

\bibitem{krause2023surface}
V.~Krause, E.~Kunze, and A.~Voigt.
\newblock A surface finite element method for the navier-stokes equations on
  evolving surfaces.
\newblock {\em Proc. Appl. Math. Mech.}, 23:e202300014, 2023.

\bibitem{art:Krause22}
V.~Krause and A.~Voigt.
\newblock A numerical approach for fluid deformable surfaces with conserved
  enclosed volume.
\newblock {\em J. Comput. Phys.}, 486:112097, 2023.

\bibitem{C9SM01902B}
Remy Kusters, Camille Simon, Rogério Lopes Dos~Santos, Valentina Caorsi,
  Sangsong Wu, Jean-Francois Joanny, Pierre Sens, and Cecile Sykes.
\newblock Actin shells control buckling and wrinkling of biomembranes.
\newblock {\em Soft Matter}, 15:9647--9653, 2019.

\bibitem{le2012buckling}
Marie Le~Merrer, David Qu{\'e}r{\'e}, and Christophe Clanet.
\newblock Buckling of viscous filaments of a fluid under compression stresses.
\newblock {\em Phys. Rev. Lett.}, 109(6):064502, 2012.

\bibitem{mongera2018fluid}
Alessandro Mongera, Payam Rowghanian, Hannah~J Gustafson, Elijah Shelton,
  David~A Kealhofer, Emmet~K Carn, Friedhelm Serwane, Adam~A Lucio, James
  Giammona, and Otger Camp{\`a}s.
\newblock A fluid-to-solid jamming transition underlies vertebrate body axis
  elongation.
\newblock {\em Nature}, 561:401--405, 2018.

\bibitem{nayyar2014stretch}
Vishal Nayyar, K~Ravi-Chandar, and Rui Huang.
\newblock Stretch-induced wrinkling of polyethylene thin sheets: Experiments
  and modeling.
\newblock {\em Int J Solids Struct}, 51(9):1847--1858, 2014.

\bibitem{nestler2019finite}
Michael Nestler, Ingo Nitschke, and Axel Voigt.
\newblock A finite element approach for vector-and tensor-valued surface pdes.
\newblock {\em J. Comput. Phys.}, 389:48--61, 2019.

\bibitem{Nitschke2022GoI}
I.~Nitschke, S.~Sadik, and A.~Voigt.
\newblock Tangential tensor fields on deformable surfaces -- {H}ow to derive
  consistent {$L^2$}-gradient flows.
\newblock {\em arXiv preprint arXiv:2209.13272}, 2022.

\bibitem{olshanskii2023}
M.~Olshanskii.
\newblock On equilibrium states of fluid membranes.
\newblock {\em Phys. Fluids}, 35:062111, 2023.

\bibitem{oratis2020new}
Alexandros~T Oratis, John~WM Bush, Howard~A Stone, and James~C Bird.
\newblock A new wrinkle on liquid sheets: Turning the mechanism of viscous
  bubble collapse upside down.
\newblock {\em Science}, 369(6504):685--688, 2020.

\bibitem{praetorius2020dunecurvedgrid}
S.~Praetorius and F.~Stenger.
\newblock {DUNE-CurvedGrid--A DUNE module for surface parametrization}.
\newblock {\em Arch. Num. Software}, pages 1--22, 2020.

\bibitem{puntel2011wrinkling}
Eric Puntel, Luca Deseri, and Eliot Fried.
\newblock Wrinkling of a stretched thin sheet.
\newblock {\em J. Elast.}, 105:137--170, 2011.

\bibitem{R2024}
Arnold Reusken.
\newblock Analysis of the taylor-hood surface finite element method for the
  surface stokes equation.
\newblock {\em arXiv preprint arXiv:2401.03561}, 2024.

\bibitem{reuther2020numerical}
Sebastian Reuther, Ingo Nitschke, and Axel Voigt.
\newblock A numerical approach for fluid deformable surfaces.
\newblock {\em J. Fluid Mech.}, 900:900, 2020.

\bibitem{SD_PNAS_1975}
P.~G. Saffman and M.~Delbr\"uck.
\newblock Brownian motion in biological membranes.
\newblock {\em Proc. Natl. Acad. Sci. USA}, 72:3111--3113, 1975.

\bibitem{Salvalaglio_2020}
Marco Salvalaglio, Mohammed Bouabdellaoui, Monica Bollani, Abdennacer Benali,
  Luc Favre, Jean-Benoit Claude, Jerome Wenger, Pietro de~Anna, Francesca
  Intonti, Axel Voigt, and Marco Abbarchi.
\newblock Hyperuniform monocrystalline structures by spinodal solid-state
  dewetting.
\newblock {\em Phys. Rev. Lett.}, 125(12):126101, 2020.

\bibitem{SanderDune2020}
O.~Sander.
\newblock {\em DUNE --- The Distributed and Unified Numerics Environment}.
\newblock Springer Cham, 2020.

\bibitem{savva2009viscous}
Nikos Savva and John~WM Bush.
\newblock Viscous sheet retraction.
\newblock {\em J. Fluid Mech.}, 626:211--240, 2009.

\bibitem{Seifert_AP_1997}
U.~Seifert.
\newblock Configurations of fluid membranes and vesicles.
\newblock {\em Adv. Phys.}, 46:13--137, 1997.

\bibitem{silveira2000rippling}
Rava~da Silveira, Sahraoui Chaieb, and L~Mahadevan.
\newblock Rippling instability of a collapsing bubble.
\newblock {\em Science}, 287(5457):1468--1471, 2000.

\bibitem{stoop2015curvature}
Norbert Stoop, Romain Lagrange, Denis Terwagne, Pedro~M Reis, and J{\"o}rn
  Dunkel.
\newblock Curvature-induced symmetry breaking determines elastic surface
  patterns.
\newblock {\em Nat. Rev. Mat.}, 14(3):337--342, 2015.

\bibitem{taylor2014spatial}
Michael Taylor, Katia Bertoldi, and David~J Steigmann.
\newblock Spatial resolution of wrinkle patterns in thin elastic sheets at
  finite strain.
\newblock {\em J. Mech. Phys. Solids}, 62:163--180, 2014.

\bibitem{tobasco2022exact}
Ian Tobasco, Yousra Timounay, Desislava Todorova, Graham~C Leggat, Joseph~D
  Paulsen, and Eleni Katifori.
\newblock Exact solutions for the wrinkle patterns of confined elastic shells.
\newblock {\em Nat. Phys.}, 18(9):1099--1104, 2022.

\bibitem{art:torres-sanchez_millan_arroyo_2019}
A.~Torres-S\'anchez, D.~Mill\'an, and M.~Arroyo.
\newblock Modelling fluid deformable surfaces with an emphasis on biological
  interfaces.
\newblock {\em J. Fluid Mech.}, 872:218--271, 2019.

\bibitem{toshniwal2021isogeometric}
Deepesh Toshniwal and Thomas~JR Hughes.
\newblock Isogeometric discrete differential forms: Non-uniform degrees,
  b{\'e}zier extraction, polar splines and flows on surfaces.
\newblock {\em Comput. Methods Appl. Mech. Eng.}, 376:113576, 2021.

\bibitem{vella2011wrinkling}
Dominic Vella, Amin Ajdari, Ashkan Vaziri, and Arezki Boudaoud.
\newblock Wrinkling of pressurized elastic shells.
\newblock {\em Phys. Rev. Lett.}, 107(17):174301, 2011.

\bibitem{vella2015indentation}
Dominic Vella, Jiangshui Huang, Narayanan Menon, Thomas~P Russell, and Benny
  Davidovitch.
\newblock Indentation of ultrathin elastic films and the emergence of
  asymptotic isometry.
\newblock {\em Phys. Rev. Lett.}, 114(1):014301, 2015.

\bibitem{vey2007amdis}
Simon Vey and Axel Voigt.
\newblock Amdis: adaptive multidimensional simulations.
\newblock {\em Comput. Vis. Sci.}, 10(1):57--67, 2007.

\bibitem{witkowski2015software}
T.~Witkowski, S.~Ling, S.~Praetorius, and A.~Voigt.
\newblock Software concepts and numerical algorithms for a scalable adaptive
  parallel finite element method.
\newblock {\em Adv. Comput. Math.}, 41:1145--1177, 2015.

\end{thebibliography}
\bibliographystyle{plain}

\end{document}